\documentclass[aps,prl,notitlepage,twocolumn,superscriptaddress,longbibliography,nofootinbib]{revtex4-1}
\pdfoutput=1
\usepackage[utf8]{inputenc}
\usepackage{amsmath,amsthm,amssymb,amsfonts}
\usepackage{scalerel}
\usepackage{mathtools}
\usepackage{graphicx}
\usepackage{subfigure}
\usepackage[normalem]{ulem}
\usepackage[colorlinks = true,
            linkcolor = blue,
            urlcolor  = blue,
            citecolor = blue,
            anchorcolor = blue]{hyperref}
\usepackage{mathrsfs}
\usepackage{bbold}
\usepackage{units}
\allowdisplaybreaks
\usepackage{bm}
\usepackage{braket}
\usepackage{makecell}
\usepackage[nameinlink]{cleveref} 
\usepackage{upgreek}
\usepackage{blindtext}
\usepackage{verbatim}
\usepackage{algorithm}
\usepackage[noend]{algpseudocode}
\makeatother

\usepackage[dvipsnames]{xcolor}
\usepackage{float}
\usepackage[bb=boondox]{mathalfa}
\usepackage{array}
\usepackage{multirow}
\usepackage{tabularx}
\usepackage{float}
\usepackage{dcolumn}
\usepackage{slashed}
\usepackage{braket}
\usepackage{verbatim}
\usepackage{multirow}
\usepackage{resizegather}
\usepackage{titlesec}
\usepackage[toc,page]{appendix}
\usepackage[export]{adjustbox}
\definecolor{teal}{RGB}{0, 158, 115} 
\definecolor{morange}{RGB}{255, 127, 0}

\DeclareMathOperator{\Tr}{Tr}
\usepackage{xcolor}
\usepackage{ulem}
\DeclareRobustCommand{\erase}{\bgroup\markoverwith{\textcolor{red}{\rule[.5ex]{2pt}{2pt}}}\ULon}

\begin{document}

\title{{Entanglement production in the Sachdev--Ye--Kitaev Model and its variants}}

\author{Tanay Pathak\,\,\href{https://orcid.org/0000-0003-0419-2583}
{\includegraphics[scale=0.05]{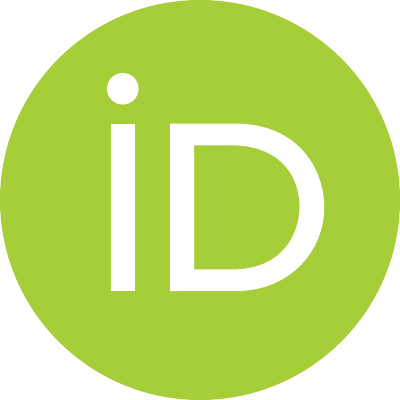}}}
\email{pathak.tanay.4s@kyoto-u.ac.jp}
\affiliation{Center for Gravitational Physics and Quantum Information, Yukawa Institute for Theoretical Physics,\\ Kyoto University, Kitashirakawa Oiwakecho, Sakyo-ku, Kyoto 606-8502, Japan}
\affiliation{Department of Physics, Kyoto University, Kitashirakawa Oiwakecho, Sakyo-ku, Kyoto 606-8502, Japan}
\author{Masaki Tezuka\,\,\href{https://orcid.org/0000-0001-7877-0839}
{\includegraphics[scale=0.05]{orcidid.pdf}}\,}
\email{tezuka@scphys.kyoto-u.ac.jp}
\affiliation{Department of Physics, Kyoto University, Kitashirakawa Oiwakecho, Sakyo-ku, Kyoto 606-8502, Japan}

\begin{abstract}
Understanding how quantum chaotic systems generate entanglement can provide insight into their microscopic chaotic dynamics and can help distinguish between different classes of chaotic behavior. Using von Neumann entanglement entropy, we study a nonentangled state evolved under three variants of the Sachdev--Ye--Kitaev (SYK) model with a finite number of Majorana fermions $N$. All the variants exhibit linear entanglement growth at early times, which at late times saturates to a universal value consistent with random matrix theory (RMT), but their growth rates differ. We interpret this as a large-$N$ effect, arising from the enhanced non-locality of fermionic operators in SYK and binary SYK, absent in spin operators of the spin-SYK model. Numerically, we find that these differences emerge gradually with increasing $N$. Although all variants are quantum chaotic, their entanglement dynamics reflect varying scrambling rates and indicate that the entanglement production rate serves as a fine-grained probe of scrambling beyond conventional measures. To probe its effect on thermalization properties of these models, we study the two-point autocorrelation function, finding no differences between the SYK variants, but deviations from RMT predictions for $N \geq 24$, particularly near the crossover from exponential decay to saturation regime.
\end{abstract}
~~~~~~~~~~~~Report Number: YITP-25-99
\maketitle

\textit{Introduction:} The Sachdev--Ye--Kitaev (SYK) model is a model of $N$ fermions with $q$-body all-to-all random interactions \cite{PhysRevLett.70.3339,kitaev1998,kitaev2015}. Given its simplicity along with rich physics, it has emerged as a paradigmatic model of quantum chaos and holography recently \cite{Maldacena:2016hyu,Garcia-Garcia:2016mno,Cotler:2016fpe,Krishnan_2018,Sarosi_2018,Maldacena:2019ufo,Trunin_2021,RevModPhys.94.035004,bousso2022snowmasswhitepaperquantum,faulkner2022snowmasswhitepaperquantum,catterall2022reportsnowmass2021theory}. 
There are proposals to realize it experimentally such as using Rydberg atoms as quantum simulators for the SYK model \cite{Schuster_2022} or recently the possibility to observe SYK physics at low temperature and in high magnetic field using graphene quantum dots \cite{Anderson_2024}. There are other proposed realization of the model such as in quantum computer simulations\cite{Danshita_2017,Garcia-Alvarez:2016wem,Franz:2018cqi,Luo_2019,jafferis2022traversable,kobrin2025experiments,Jafferis2025,Asaduzzaman:2023wtd}.
Efforts have been made to further simplify the model 
for a more feasible experimental realization. Variants of interest 
are the sparse SYK model \cite{Xu:2020shn,Caceres:2021nsa, PhysRevD.103.106002,Anegawa_2023,Caceres:2023yoj,orman2024quantumchaossparsesyk, Tezuka:2022mrr} and the spin SYK model \cite{Swingle:PhysRevB.109.094206,Hanada_2024}. Various other variants have been studied in the literature for their interesting properties in high energy as well as condensed matter physics \cite{French:1970ztu,Bohigas:1971vpj,PhysRevX.5.041025,PhysRevB.105.075117,PhysRevB.105.235131,PhysRevB.95.155131,Gu_2020,PhysRevResearch.2.033025,PhysRevB.94.035135,PhysRevB.100.155128,Gross_2017,PhysRevD.95.026009,Li_2017,PhysRevLett.124.244101,Gates_2021,PhysRevX.12.021040,PhysRevLett.130.010401,Nandy:2024wwv,maldacena2018eternal,Jia_2022,Berkooz_2017,Gu_2017,Berkooz:2018jqr,Berkooz:2018qkz,Ozaki:PhysRevResearch.7.013092,PhysRevB.103.195108,PhysRevLett.105.151602,PhysRevLett.119.216601}. 

In this paper, we focus on three variants: the sparse SYK model, the binary sparse SYK model and the sparse spin SYK model.
For deciding whether they are suitable targets for practical implementation, a natural question to ask is: \emph{To what extent are these models similar(different) to(from) each other?} 
Some comparisons based on information recovery and finite-$N$ spectrum of these variants had been carried out in \cite{Monteiro_2021,Tezuka:2022mrr,Nakata_2024,Hanada_2024}.
In this work we attempt the first systematic effort to quantitatively compare the SYK, sparse variants of SYK, and RMT for their (dynamical) chaotic properties using the entanglement production process. The results presented here thus also serve as a sensitive test to compare the result of SYK model with corresponding random matrices.
The entanglement entropy has already been well studied for its relationship with quantum chaos \cite{RevModPhys.80.517,Wang_2004,Santos_2004,Bandyopadhyay_2004, Karthik_2007,Dogra_2019,Trail_2008,PhysRevLett.119.220603, PhysRevLett.125.180604}. It has also been observed that given a less entangled(unentangled) state evolving under the action of a given quantum chaotic Hamiltonian, its entanglement is substantially enhanced \cite{ Zurek:1995jd,PhysRevLett.80.5524,PhysRevE.60.1542,PhysRevE.64.036207,Zanardi:2000zz, Zanardi_2001,Scott_2003,Scott_2004, bandyopadhyay2005ep,Abreu_2006}. For strongly chaotic system the entanglement production saturates at late times \cite{PhysRevE.64.036207,Bandyopadhyay_2004} which is a statistical property and based on random matrix theory (RMT) modeling a statistical bound on entanglement entropy was proposed in \cite{Bandyopadhyay_2002}. The entanglement production can thus be used as a criterion of dynamical chaos in quantum chaotic systems\cite{LAHIRI20036,lahiri2003dynamical,Gharibyan_2019}. We use the entanglement production rate as a metric to compare the three variants of the SYK models and quantify their chaotic nature. 

\textit{Random Matrix Theory:} We briefly recapitulate the bound on the entanglement entropy based on random matrix theory modeling as discussed in \cite{PhysRevE.64.036207,Bandyopadhyay_2002} for completeness. Consider a bi-partition of the Hilbert space, $\mathcal{H} =  \mathcal{H}_{A} \bigotimes \mathcal{H}_{B} $ of dimension $\mathcal{D}$. Without loss of generality we consider, $\dim(\mathcal{H}_{A}) = \mathcal{N} \leq$  $\dim(\mathcal{H}_{B}) = \mathcal{M}$, $\mathcal{D}= \mathcal{N} \times \mathcal{M}$. The reduced density matrix constructed from an eigenstate of random matrices (by tracing out, say system $B$) belong to ensemble of trace restricted Wishart matrices \cite{PhysRevE.64.036207,Bandyopadhyay_2002}; $ \rho_{A,B} = \frac{G^{\dagger}G}{\Tr(G^{\dagger}G)}$,
where $G$ is an unstructured $\mathcal{N} \times \mathcal{N}$ Gaussian random matrix. The average density of states of ensemble of Wishart matrices is given by the Marchenko--Pastur distribution \cite{mehta2004random} and using it a bound on the average entanglement entropy was proposed in \cite{Bandyopadhyay_2002}, which can be compactly written as \cite{Tomsovic_2018} (see also \cite{supp}), as follows
\begin{equation}\label{eq:eesimple}
    \braket{S_E} \cong  \ln(\mathcal{N}) - \frac{1}{2 Q};\quad Q= \frac{\mathcal{M}}{\mathcal{ N}}
\end{equation}
which corresponds to Page's result for the average EE of a random pure state, for $1 \ll  \mathcal{N} < \mathcal{M} $ \cite{Page:1993df,Sen_1996}. If we now start with an initial state and let it evolve in time using a quantum chaotic Hamiltonian then at late times the EE will saturate the statistical bound, Eq.\eqref{eq:eesimple}. For the extreme case where the initial state is a product state, the EE will be zero at the start and will eventually grow to saturate the bound at late times. For the other case where the initial state is a maximally entangled state, for which the EE is already maximal (and also greater than the above bound), the evolution in time will lead to an initial \emph{decrease} of EE which then ultimately saturates the bound at late times \cite{Bandyopadhyay_2002}. We verify these results numerically for the case of GOE and GUE matrices (see Supplemental material \cite{supp}).

\textit{The models:}
We mainly focus on the three variants of the SYK model which were proposed for their feasibility of simulation in quantum computers (see \cite{PhysRevLett.111.127205} for study in spin models). The SYK model preserves parity, therefore the Hamiltonian can be written in block diagonal form, with two blocks \cite{Krishnan_2018}. For each of these models, care is also taken to ensure that the variance of eigenvalues is the same (see Supplemental material for details on numerical methods \cite{supp}). This is important to ensure proper comparison of all the \emph{relevant time scales} among the three models. Furthermore, the product state ($\psi_P$) that we consider throughout the paper is \cite{note1}
\begin{align}
    \Psi_P &= \bigotimes_{i=1}^n \frac{1}{\sqrt{2}}(\ket{0}+\ket{1}). 
\end{align}

1. \textit{The sparse SYK model:}
\begin{figure*}[t]
    \centering
    \includegraphics[width=  0.9\linewidth]{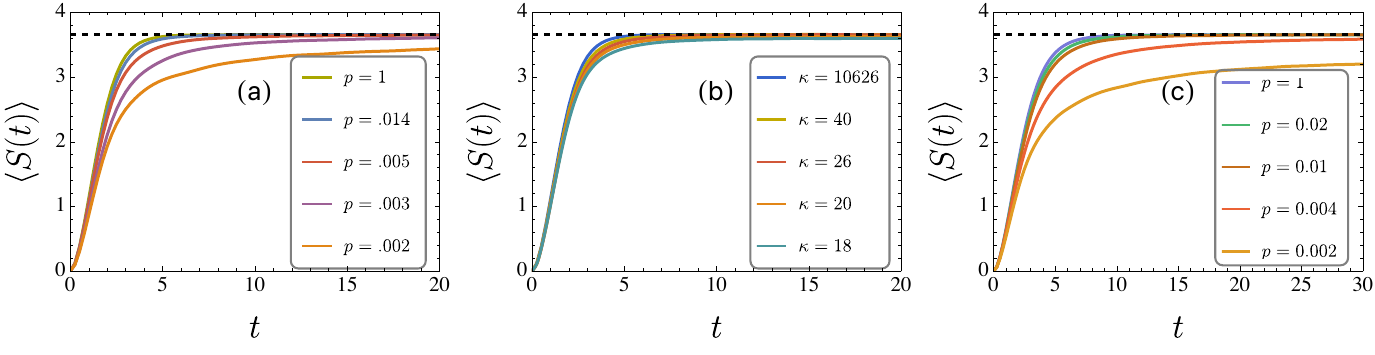}
 \onecolumngrid \caption{(a), (b), (c) Evolution of EE of a product state with time, for SYK, binary SYK and the spin SYK model respectively. We do not resolve symmetry but it can be shown that the qualitative behavior remains the same (see \cite{supp} for further detials on symmetry). The horizontal black line corresponds to the analytical result given by Eq. \eqref{eq:eesimple}. The horizontal black line corresponds to the analytical result given by Eq. \eqref{eq:eesimple}. We take  $N =24$ for SYK and SYK$_b$ $N =12$ for SYK$_s$. Averaging is done over 50 realizations. The results are only shows for sparseness parameter values for which the different curves can be distinguished properly (see \cite{supp} for result of other intermediate values of sparse parameter $p$ and $\kappa$).
    }
    \label{fig:syk_ee}
\end{figure*}
We start with the usual SYK model \cite{Maldacena:2016hyu} which is a model of $q$-fermion interactions whose Hamiltonian is given by 
\begin{align}
    H = \sqrt{\frac{(q-1)!}{N^{q-1}}} \sum_{1 \leq i_{1} < \cdots i_{q} \leq N } J_{i_{1}i_{2}\cdots i_{q}}\psi_{i_1}\psi_{i_2} \cdots \psi_{i_q},
\end{align}
where $\psi_{i}$ are the Fermionic operators which satisfy the Clifford algebra: $ \{\psi_{i},\psi_{j}\}= \delta_{ij}$
and $J_{i_{1}i_{2}\cdots i_{q}}$ are standard Gaussian random variables with probability distribution function 
$P(J_{i_{1} \cdots i_{q}}) = \frac{1}{\sqrt{2 \pi}}e^{-J_{i_{1} \cdots i_{q}}^{2}/2}$. We specialize to the $q=4$ case (SYK$_4$, which we call SYK for brevity). Furthermore, we consider a simple generalization of the SYK model with the aim that the features of Hamiltonian given by Eq. \eqref{eq:syk4ham} remains intact and at the same time with lesser complexity is the sparse SYK model \cite{Xu:2020shn,Caceres:2021nsa, PhysRevD.103.106002,Anegawa_2023,Caceres:2023yoj,orman2024quantumchaossparsesyk}  which is defined as follows
\begin{equation}\label{eq:syk4ham}
     H = \sqrt{\frac{6}{p N^{3}}} \sum_{1 \leq i<j<k<l \leq N } x_{ijkl}J_{ijkl}\psi_{i}\psi_{j} \psi_{k} \psi_{l},
\end{equation}
where $x_{ijkl}$ is randomly chosen to be 1 or 0 with probability $p$ and $1- p$, respectively, and $J_{i_{1}i_{2}i_{3} i_{4}}$ are standard Gaussian random variables.
The model is then equivalent to the deletion of some terms in the dense SYK model Hamiltonian(implying no deletion of terms i.e. $p=1$).
~~~~~~~~~~~~~~~~~~~~~~~~~~~~~~~~~~~~~~~~~~~~~~~~~~~~~~~
\noindent 2. \textit{The Binary SYK model:} Another important variant of the SYK model is binary sparse SYK model (SYK$_b$) where the couplings of the model are nonzero with probability $p$ and zero with probability $1-p$. The non-zero couplings are either $+1$ and $-1$ with probability $p/2$ each. This model is a simplification of the original model, and as has been pointed out in \cite{Tezuka:2022mrr} it is an improvement and matches with the RMT predictions better as compared to the corresponding sparse SYK model. For the numerical purpose, we follow \cite{Tezuka:2022mrr} and keep the number of non-zero couplings as fixed, $\kappa$. The number of couplings $J_{i_1 i_2 i_3 i_4}$ that take values $\pm 1$, is $\kappa/2$ when $\kappa$ is even, and $(\kappa \pm 1)/2$ when $\kappa$ is odd. The remaining, ${{N}\choose{4}} - \kappa$, couplings are 0. From this we can obtain $p$ as $p= \frac{\kappa}{{{N}\choose{4}}}$. The Hamiltonian is given by 
\begin{align}
    H = \sqrt{\frac{6}{p N^{3}}} \sum_{1 \leq i_{1} < i_{2}<i_{3}< i_{4} \leq N } J_{i_{1}i_{2} i_{3} i_{4}}\psi_{i_1}\psi_{i_2}\psi_{i_3}\psi_{i_4},
\end{align}
\noindent 3. \textit{The spin-SYK model:} The spin-SYK (SYK$_s$) is a variant where the Majorana fermions in the SYK model are replaced by Pauli spin operators \cite{Hanada_2024}. The spin analogue of fermionic operators, $\hat{O}_{a}$, are defined as: $
    \hat{O}_{2j-1}= \hat{\sigma}_{j,x}$,$ \hat{O}_{2j}= \hat{\sigma}_{j,y}$ $; j= 1,2,3,...N.
$. Here, $\sigma_{i,k}= \hat{I}_{i-1} \otimes \sigma_{i,k} \otimes \hat{I}_{N-i}$ and $\hat{I}_{l}$ denotes the $2^{l}$-dimensional identity operator. Operators on the same spin anti-commute while operators on different spins commute. 
The sparse spin SYK Hamiltonian is then given as
\begin{align} \label{eq:ssyk4ham}
     H_{s} = \sqrt{\frac{6}{p (2 N)^{3}}}\sum_{1 \leq i<j<k<l \leq 2 N } &x_{ijkl}\,J_{ijkl}\, i^{\eta_{ijkl}} \hat{O}_{i}\hat{O}_{j} \hat{O}_{k} \hat{O}_{l},
\end{align}
where $J_{ijkl}$ are standard Gaussian random variables, $x_{ijkl}$ is $1$ with probability $p$ and $0$ with probability $1-p$, similar to the case of the sparse SYK model. $p=1$ corresponds to the full dense spin SYK model. $\eta_{ijkl}$ is the number of spins whose both $x$ and $y$ components appear in $(i, j, k, l)$ and the factor $i^{\eta_{ijkl}}$ ensures hermiticity of the Hamiltonian. Note that the variance of eigenvalues remains the same as in the previous two variants.

\noindent\textit{Results: Entangling rate--}
First we show the general observed behavior for the three variants for different degrees of sparseness in Fig. \eqref{fig:syk_ee}. For all the three variants we find a initial \emph{linear} increase in entropy which then saturates at later time to a value given by Eq. \eqref{eq:eesimple}. We further note that after a certain degree of sparseness, the entanglement production is not enough to saturate the bound. This critical value after which the entanglement entropy does not reach the bound (numerically checked till $t= 30$ everywhere) is found to be: \emph{$p_{c}= 0.004$} for SYK, \emph{$\kappa_{c}= 22$} ($p_{c} \approx 0.002$) for SYK$_b$ and $p_{c}= 0.01$ for SYK$_s$.  This scenario can be attributed to the chaotic to integrable transition of the sparse SYK model and variants as the value of $p$ ($\kappa$) decreases (increase of sparsity) \cite{PhysRevD.103.106002,orman2024quantumchaossparsesyk,Tezuka:2022mrr}. It is also to be noticed that the critical sparsity at which the breakdown occurs is lowest for SYK$_{b}$ while it is highest for SYK$_{s}$.

\begin{figure}[h]
    \centering
    \includegraphics[width= \linewidth]{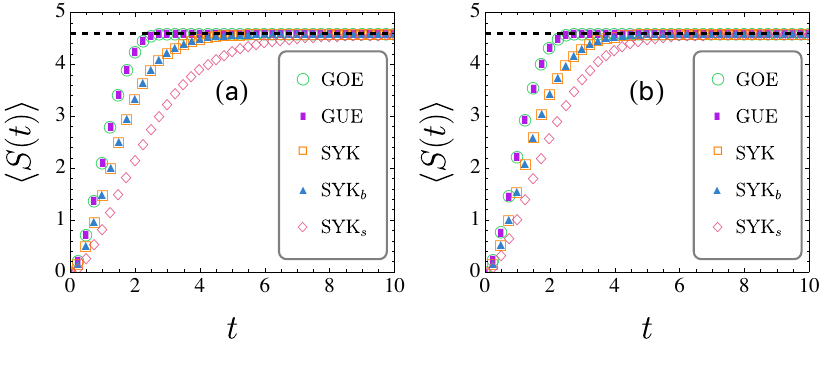}
    \caption{(a)The evolution of the EE for initial product state using (dense; $p=1$) SYK and SYK$_b$ with $N=30$ fermions (no symmetry resolution) and SYK$_s$ with $N=15$ spins. Also shown are the results for random matrices of GOE and GUE type and dimensions $2^{15}$. We consider 50 realization for each of them. Slopes of best fit line in the linear regime (observed for $t \geq 0.5$ for all the models) are-- GOE:$2.65 \pm 0.1$, GUE:$2.65 \pm 0.1$, SYK:$2.05 \pm 0.03$, SYK$_b$:$2.05 \pm 0.03$, and SYK$_s$:$1.51 \pm 0.03$. (b) Same as (a), but with SYK and SYK$_b$ with $N=32$ fermions and SYK$_s$ with $N=15$ spin, and considering only the odd parity block. Slopes of best fit line in the linear regime (observed for $t \geq 0.5$ for all the models) are--GOE:$2.60 \pm 0.08$, GUE:$2.60 \pm 0.08$, SYK:$1.99 \pm 0.02$, SYK$_b$:$1.99 \pm 0.02$, SYK$_s$:$1.26 \pm 0.03$. In both cases, SYK models show systematic deviation from random matrix predictions.}
    \label{fig:sykall_ee}
\end{figure}

Next, we contrast the behavior of the dense versions of the three variants and compare the entanglement production rate of these models with the GOE and GUE-type random matrices \cite{supp}. With initial product state, 
the entanglement entropy grows linearly and then saturates at the universal value as shown in Fig. \ref{fig:sykall_ee}.
GOE and GUE type random matrices have the same rate while all the variants of the SYK model show deviations from the (Gaussian) random matrix behavior. Furthermore, SYK and SYK$_{b}$ have the same slope in the linear growth regime that differs from the SYK$_s$ model.
Thus we find that all the variants form a clear hierarchy based on their entangling rate: \emph{SYK$_b$ $ \approx $ SYK $>$ SYK$_s$} and also show deviations from the (Gaussian) random matrix behavior.  Note that if we take sparseness into account we find that the SYK$_b$ retains these features for much smaller values of sparseness parameter $p$ as compared to SYK and SYK$_s$. 
\begin{figure}[ht]
    \centering
    \includegraphics[width=\linewidth]{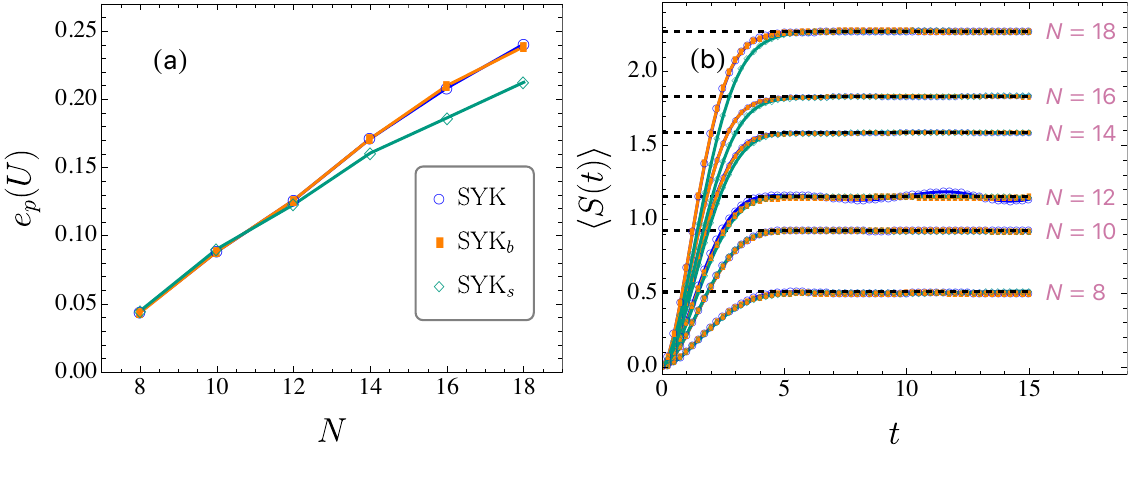}
    \caption{(a) Entangling power (parity resolved) calculated as a function of $N$.(b) Evolution of average EE of initial product state, $\psi_{p}$, for SYK model (parity resolved). We consider $N$=$ 8(10^4)$,$ 10(5000)$, $12(2000)$, $14(1000)$, $16 (500)$$,18 (250),$ number of Majorana fermions(correspondingly we consider $N/2$ spins for spin SYK), where values in the bracket denotes the number of Hamiltonian realization considered. For $e_{p}(U)$ both the Hamiltonian and the initial state are randomly generated in each iteration.  Both $e_{p}(U)$ and EE are calculated by partially tracing out $2^{2}$ dimensional subspace for $N= 8$$,10,12$; $2^{3}$ dimensional subspace for $N= 14,16$ and $2^{4}$ dimensional subspace for $N= 18$. The dashed line denotes the thermal value \cite{Page:1993df}(we cannot directly use Eq.\eqref{eq:eesimple} as the subsystem sizes are too small for it to be valid). Observe the gradual deviations of the entangling power and the rate of growth of EE for SYK$_s$ as $N$ increases.}
    \label{fig:syksmalln_ee}
\end{figure}
Based on the detailed numerical results presented here and previous studies on the SYK model and its variants we conclude that SYK$_s$ is slightly less (dynamically) chaotic than the other two. The entanglement under its action eventually reaches the thermal value, though slowly, suggesting different 
microscopic
thermalization properties. The origin of this behavior can be attributed to the locality in SYK$_s$ which is absent in the other two models once the Majorana operator and spin operators of the model are written as Pauli string \cite{Hanada_2024}. 
To confirm this assertion, we study the effect of $N$ on the evolution of EE for small values of $N$ where the locality should not play a major role and hence all the models should behave in the same manner. As $N$ increases SYK and SYK$_b$ have more and more operators which have non-local Pauli strings, while SYK$_s$ still has local strings. Hence, strictly speaking, the deviation of SYK$_s$ results from SYK and SYK$_{b}$ should be a large $N$ effect and not observed for small $N$.
Furthermore, to remove any dependence on the initial state we also consider the entangling power, which is defined as follows \cite{Zanardi:2000zz},
\begin{equation}
   e_{p}(U)= \overline{E_{l}(U\ket{\psi})},
\end{equation}
where $E_{l}= 1- \mathrm{Tr}(\rho_{A}^{2})$ denotes the linear entropy of subsystem $A$, the bar denotes the average over all produce state $\psi= \psi_{1}\otimes \psi_{2}$ and $U= e^{- i H t}$ is the time evolution operator. In Fig. \ref{fig:syksmalln_ee}(a), we show the entangling power for different values of $N= 8, \ldots, 18$. We observe no discernible differences between the three variants up to $N=12$. From $N=14$ the deviations become prominent as $N$ increases further. We also show the results for the entanglement production with initial product state and different values of $N$ in Fig. \ref{fig:syksmalln_ee}(b) which corroborate the observations for entangling power. This thus confirms our assertion.

\begin{figure}[h]
    \centering
    \includegraphics[width=  \linewidth]{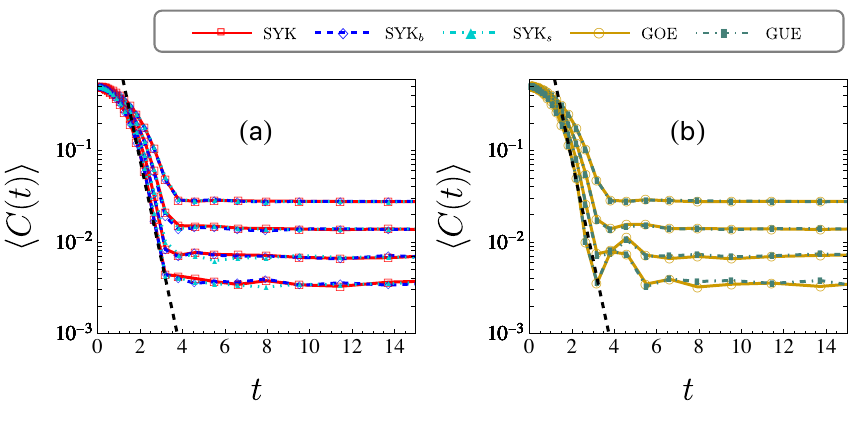}
    \caption{(a) Autocorrelation function for Majorana operator $\hat{O}= \psi_{1}$, for different kinds of SYK model. The overlapping legends are for SYK, SYK$_b$ and SYK$_s$ for $N= 18 (5000), 22 (1000), 26(200),30(100)$ of Majorana fermions (and $N/2$ number of spins for SYK$_s$). The number in the bracket denotes the number of Hamiltonian realizations. We only consider odd parity sector. (b) Autocorrelation function for corresponding random matrices of GOE and GUE type of dimensions $2^{N/2-1}$. Black dashed curve represent the best fit curve; $a e^{-b x} $; $b= 2.49 \pm 0.23 $ ,  in the exponential decay regime obtained using the data for $2^{14}$ dimensional random matrices.}
    \label{fig:acfall}
\end{figure}

\noindent\textit{Two-point correlation function--}With the differences in the entangling rate, it is thus natural to study its effects on the thermalization and operator growth properties of the model. As a first step towards studying thermalization properties, we study two-point autocorrelation function for these models. Choosing the initial state as the product state, $\psi_{p}$, and Majorana operator $\psi_{1}$, we study the two-point autocorrelation $\braket{C(t)}= \big \langle \braket{\Psi_{p}|\psi_{1}(t)\psi_{1}|\Psi_{p}}\big \rangle$, where $\big \langle \boldsymbol{\cdot} \big \rangle $ denotes the ensemble average over different Hamiltonian realizations. The results for the odd parity sector of the SYK Hamiltonian are shown in Fig. \eqref{fig:acfall}. 
We observe that the autocorrelation function has the characteristics features that of chaotic systems: an \emph{exponential} decay regime at early times and a saturation value at late times that decreases with $N$ (see also \cite{supp} for supporting results). 
Though we do not find differences between the three variants, we observe slight deviations from RMT for $N \geq 24$ near the transition from decay to saturation regime. These differences between RMT and the SYK model are observed to increase with $N$.

\textit{Conclusions:} In this work we study and contrast the entanglement growth of an initial unentangled state evolved under the SYK model, binary SYK (SYK$_{b}$) and spin SYK (SYK$_{s}$). Using extensive numerical analysis we show that the SYK$_{s}$ model shows the slowest entanglement growth rate as compared to others, which show identical behavior. We assert that this behavior is due to the \emph{local structure} in the SYK$_s$ model. We then  numerically confirm this assertion for small $N$ where this effect would not be significant and hence all the models show identical entanglement growth rates. We also remark that, although eigenvalue correlation features of all these models closely follow the RMT prediction (with small deviations near the tail for the spin-SYK), their entanglement production rates 
do not match the Gaussian RMT behavior. Though the production rates in SYK and SYK$_b$ are closest to the RMT values, they are always found to be slower, at least for the finite size systems considered. These results are some of the only differences from RMT found for the SYK model variants. The entanglement production rates can therefore serve as a sensitive dynamical probe, highlighting differences between physical models and their associated RMT ensemble that ignore the fine-grained details of these models. To study its effect on thermalization properties we also studied the autocorrelation function. We observe the characteristic, early time exponential decay and late time saturation behavior, typical for quantum chaotic systems, for all the variants. We find good agreement in the decay and saturation regimes, 
with small differences only near the transition regime and large $N(\geq 24)$. It is important to note that in holographic theories the butterfly velocity ($v_{B}$) that governs the rate of the growth of an operator under chaotic dynamics is related to the entanglement velocity ($v_{E}$), i.e. the rate at which the entanglement spreads \cite{Hartman:2013qma, Liu:2013iza, Liu:2013qca, Hosur:2015ylk} as $v_{E} \leq v_{B}$. This further links the entangling rate to the properties of operator growth in the systems. Since locality played a major role in the entanglement production, it will also be interesting to study other variants of the SYK model such as the recently introduced two-local modification \cite{Hanada:2025pis} within this framework. 

\emph{Acknowledgments:}
Numerical computations were performed using the computational facilities of the Yukawa Institute for Theoretical Physics. TP acknowledges the partial support of the Yukawa Research Fellowship, supported by the Yukawa Memorial Foundation and JST CREST (Grant No. JPMJCR19T2).
The work was partially supported by JST CREST (Grant No. JPMJCR24I2).
M.~T. was partially supported by the Japan Society for the Promotion of Science (JSPS) Grants-in-Aid for Scientific Research (KAKENHI) Grants No. JP21H05185 and JP25K00925.

\textit{Data availability} — The data that support the findings of this study are openly available at Zenodo at the link in Ref. \cite{dataall}.
\bibliography{references}
\clearpage
\end{document}


\title{{Supplemental Materials: Entanglement production in the Sachdev--Ye--Kitaev Model and its variants}}

\author{Tanay Pathak\,\,\href{https://orcid.org/0000-0003-0419-2583}
{\includegraphics[scale=0.05]{orcidid.pdf}}}
\email{pathak.tanay.4s@kyoto-u.ac.jp}
\affiliation{Center for Gravitational Physics and Quantum Information, Yukawa Institute for Theoretical Physics,\\ Kyoto University, Kitashirakawa Oiwakecho, Sakyo-ku, Kyoto 606-8502, Japan}
\affiliation{Department of Physics, Kyoto University, Kitashirakawa Oiwakecho, Sakyo-ku, Kyoto 606-8502, Japan}
\author{Masaki Tezuka\,\,\href{https://orcid.org/0000-0001-7877-0839}
{\includegraphics[scale=0.05]{orcidid.pdf}}\,}
\email{tezuka@scphys.kyoto-u.ac.jp}
\affiliation{Department of Physics, Kyoto University, Kitashirakawa Oiwakecho, Sakyo-ku, Kyoto 606-8502, Japan}
\maketitle

\section{Fixing the variance of eigenvalues}\label{appendix:rmt}

We now provide the method to fix the variance across all the models and random matrices considered. It is important to fix the energy scales, which we achieve by fixing the variance, to ensure proper comparison of the models. We first note a few properties of the random matrices belonging to Gaussian orthogonal ensemble (GOE) and Gaussian Unitary ensemble(GUE). 

GOE type random matrices are real Hermitian matrices with entries chosen, upto symmetry, from Gaussian distribution with variance 1 along the diagonal and variance $1/2$ along the off-diagonals. Similarly, GUE type random matrices are Hermitian matrices with complex entries (upto Hermiticity) with both real and imaginary parts chosen from Gaussian distribution with variance 1/2.

Now, using the property that for a given $\mathcal{D} \times \mathcal{D}$ dimensional matrix $H$ its trace can be written as $\mathrm{Tr}(H^2)= \sum_{m=1}^{\mathcal{D}}\sum_{m=1}^{\mathcal{D}} H_{mn}^{2}$ we can calculate the expectation value of $\mathbb{E}(\mathrm{Tr}(H^{2})$ for both ensembles. We then have 
\begin{align}
    \mathbb{E}(\mathrm{Tr}(H_{\mathrm{GOE}}^{2})) &= \sum_{i}\mathbb{E}((\lambda_{\mathrm{goe}})_{i}^{2})= \frac{\mathcal{D}(\mathcal{D}+1)}{2} \\
    \mathbb{E}(\mathrm{Tr}(H_{\mathrm{GUE}}^{2})) &=\sum_{i}\mathbb{E}((\lambda_{\mathrm{gue}})_{i}^{2})= \mathcal{D}^{2}.
\end{align}
where $\lambda_{i}$ denotes the eigenvalue of the corresponding random matrix.
For the SYK model (all the variants), with $N$ fermion and $q=4$, using the property of Majorana operator we can calculate the expectation value of trace of hamiltonian squared as 
\begin{equation}
   \mathbb{E}(\mathrm{Tr}(H_{\text{SYK}}^{2}))= \sum_{i}\mathbb{E}((\lambda_{\mathrm{\text{SYK}}})_{i}^{2})= \frac{6\ 2^{N/2} \binom{N}{4}}{2^4 N^3}
\end{equation}
Using these properties we can calculate the variance of eigenvalues, and we get the following:

\begin{equation}
    \sigma^{2}_{\mathrm{GOE}}= \frac{(\mathcal{D}+1)}{2} \quad \sigma^{2}_{\mathrm{GUE}}= \mathcal{D}, \quad \sigma^{2}_{\mathrm{SYK}}= \frac{6\ 2^{N/2} \binom{N}{4}}{2^{4+N/2} N^3}
\end{equation}
Using these properties we can easily fix the variance of the corresponding semicircle eigenvalue density. As an example of the above implementation consider we want to match the variance of GOE random matrix with SYK model having $n$ Majorana fermions\stepcounter{footnote}%
\footnotemark[\value{footnote}]%
\footnotetext[\value{footnote}]{We avoid using \texttt{N} as it is an internal command in \textsc{Mathematica}.}. In \textsc{Mathematica} one can use the command:
\begin{center}
   \texttt{RandomVariate[GaussianUnitaryMatrixDistribution[1, $2^{n/2}$]} 
\end{center}
to generate a random matrix of GOE type. We then multiply the generated matrix with the prefactor $\frac{1}{2^{n/2}} \sqrt{\frac{6\ 2^n \binom{2 n}{4}}{2^4 (2 n)^3 2^n}}$. This will fix the variance automatically. Other cases can be dealt with in a similar manner. Notice that this procedure is strictly valid for large-dimensional random matrices and one will see differences at small sizes due to the comparable finite size corrections to the semicircle eigenvalue density.

\section{Results for random matrices}\label{appendix:rmt}

\begin{figure}[h]
    \centering
    \includegraphics[width=  0.5\linewidth]{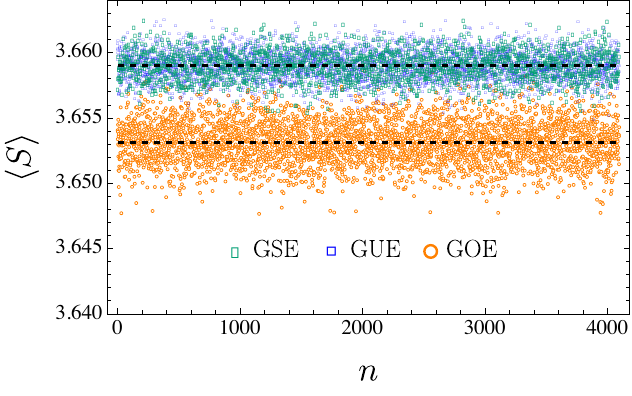}
    \caption{$\braket{S}$ of the eigenstates, indexed using $n$, of GSE, GUE and GOE type random matrices, for $l=1/2$ sub-system fraction. The horizontal black dashed lines correspond to the analytical results, for GUE (top) and GOE (bottom), available in \cite{Kumar:2011yxa}. The horizontal three different markers corresponds to the numerically obtained result for GSE, GUE and GOE respectively. We take matrices of dimensions $2^{12}$ and averaging is done over 50 realizations (for each index $n$). Observe that the results for GSE and GUE are similar, as we use the typical $2N \times 2N$ dimensional complex representation of $N \times N$ dimensional quaternionic matrices.}
    \label{fig:rmtstateee}
\end{figure}

In this appendix, we provide the numerical results for the average entanglement entropy (EE) of the eigenstates of Gaussian random matrices belong to orthogonal (GOE), unitary (GUE) and symplectic (GSE) class. The bound for average entanglement entropy for such eigenstates using RMT is given as
\begin{equation}\label{eq:eesimplesupp}
    \braket{S_E} \cong  \ln(\mathcal{N}) - \frac{\mathcal{N}}{2 \mathcal{M}}
\end{equation}
However, the above result is valid in the case of large $\mathcal{N}$ and $\mathcal{M}$ and do not account for Dyson index, $\beta$ dependence. The full results with $\beta$ dependence are obtained in\cite{Kumar:2011yxa}, which we do not write here due to their lengthy expressions. 
These results are shown in Fig. \ref{fig:rmtstateee}. The mean of the average EE over the whole spectrum (and 50 Hamiltonian realization) we obtain are: $\braket{S}_\mathrm{GSE}= 3.659$, $\braket{S}_\mathrm{GUE}= 3.659$, $\braket{S}_\mathrm{GOE}= 3.653$ for GSE, GUE and GOE respectively  while using the Eq.\eqref{eq:eesimplesupp} we obtain $\braket{S}= 3.659$. Using the full result (with full $\beta$ dependence) in \cite{Kumar:2011yxa} we have $\braket{S}_\mathrm{GSE}= 3.662$, $\braket{S}_\mathrm{GUE}= 3.659$, $\braket{S}_\mathrm{GOE}= 3.653$.

The following states are used as the product and the maximally entangled initial states:
\begin{align}
    \Psi_P &= \bigotimes_{i=1}^n \frac{1}{\sqrt{2}}(\ket{0}+\ket{1}), 
    \\
    \Psi_M &= \frac{1}{\sqrt{n}} \sum_{i=1}^{n} (\ket{i}\rangle \ket{i}\rangle),
\end{align}
where in $\Psi_{M}$, $\ket{1}\rangle = (1,0,0,\cdots,0), \ket{2}\rangle = (0,1,0,\cdots,0) $  and so on denotes a $n$-level basis state.

\begin{figure}
\centering
\begin{minipage}{.5\textwidth}
  \centering
  \includegraphics[width=\linewidth]{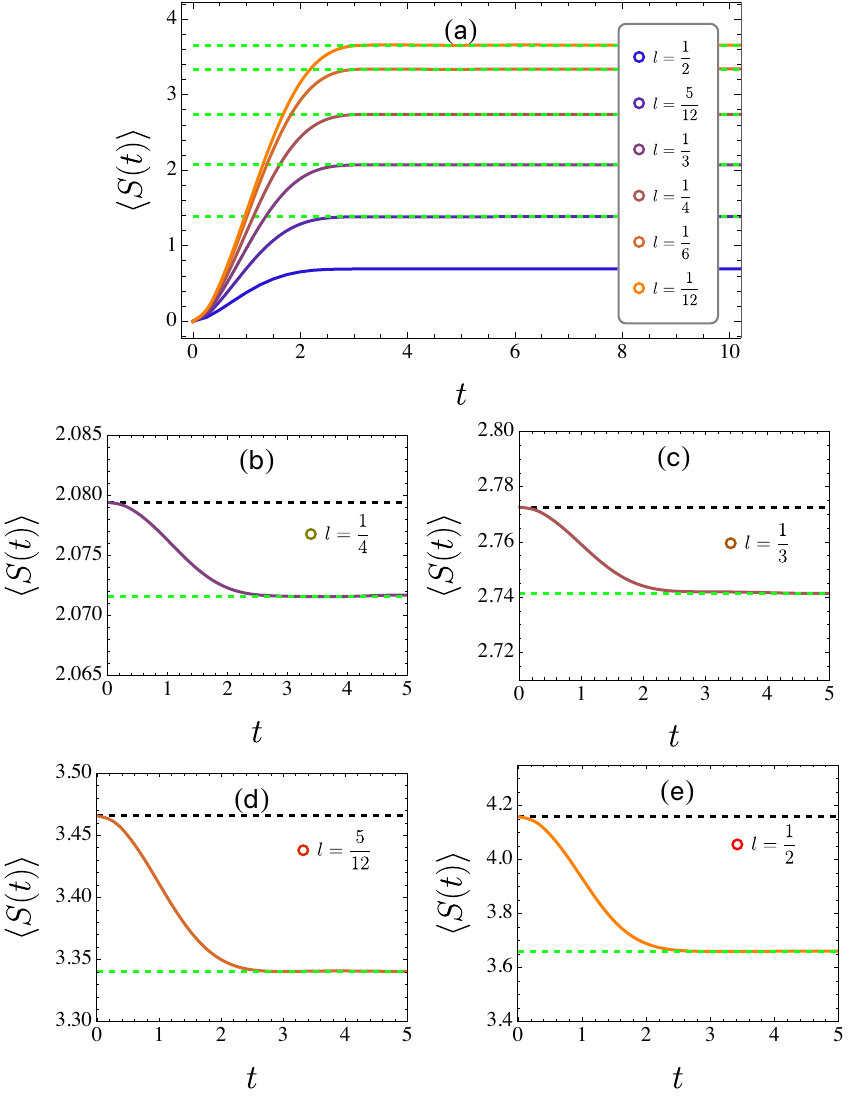}
\end{minipage}%
\begin{minipage}{.5\textwidth}
  \centering
  \includegraphics[width=\linewidth]{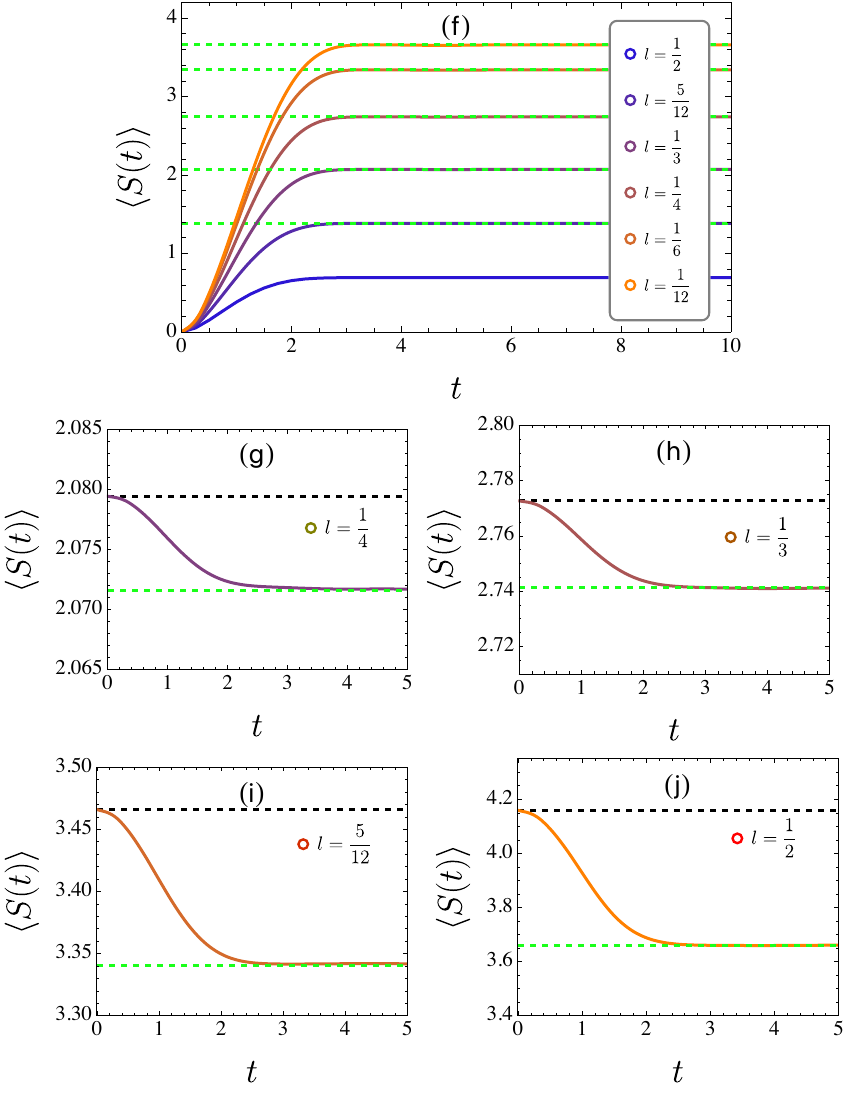}
\end{minipage}
\caption{(Left) (a) Evolution of EE for an initial product state with time, for various sub-system fractions, for the GOE case. (b), (c), (d), (e) show the EE evolution, using GOE matrices, of maximally entangled state with subsystem fraction $l= 1/4, 1/3, 5/12, 1/2$. The horizontal dashed black line corresponds to the maximal value of entropy and the horizontal  green line corresponds to the analytical result given by Eq. \eqref{eq:eesimplesupp}. (Right) (f) Evolution of EE for an initial product state with time, for various sub-system fractions, for the GUE case. (g), (h), (i), (j) show the EE evolution, using GUE matrices, of maximally entangled state with subsystem fraction $l= 1/4, 1/3, 5/12, 1/2$. The horizontal dashed black line corresponds to the maximal value of entropy and the horizontal  green line corresponds to the analytical result given by Eq. \eqref{eq:eesimplesupp}. We take $\mathcal{D} =2^{12}$ and averaging is done over 50 realizations.}\label{fig:rmt_ee}
\end{figure}

In Fig. \ref{fig:rmt_ee}(left) (a) we provide the results of the evolution of initial product state $\Psi_{P}$. Fig. \ref{fig:rmt_ee}(left) (b)-(e) show the result for the maximally entangled initial state $\Psi_{M}$ which is evolved using a GOE type random matrix of dimension $\dd =2^{12}$, for various values of subsystem fraction $l$. The averaging is done over 50 Hamiltonian realizations. We observe a good agreement between the saturation value and the theoretical limit. As was briefly discussed in the main text, we also observe that the bound is saturated for both $\Psi_{P}$ and $\Psi_{M}$. For the case of $\Psi_M$, as the initial state is already maximally entangled, the entanglement entropy first decreases (the state gets slightly disentangled) and then saturates to the theoretical value given by Eq. \eqref{eq:eesimplesupp}.  In Fig.~\ref{fig:rmt_ee} (f)-(j), the results are shown for the GUE with similar conclusions. 

\section{Derivation of the entanglement bound }\label{appnedix:derboud}

The final result obtained in this section is already stated in \cite{Tomsovic_2018} and can also be easily obtained from the result derived in \cite{Page:1993df,Sen_1996}. We however re-derive the result here again for the convenience of the readers.

Consider a bi-partition of Hilbert space $\mathcal{H} =  \mathcal{H}_{A} \bigotimes \mathcal{H}_{B} $, of dimension $\mathcal{D}$ . We have in general $dim$(\h$_A$) = $\mathcal{M}$ and $dim$(\h$_B$) = $\mathcal{N}$, $\mathcal{D}= \mathcal{M} \bigotimes \mathcal{N}  $. Consider the density matrix $\rho \in$ \h. The reduced density matrix (RDM) obtained after tracing over subsystem $A$ is given by $\rho_{B}= \Tr_{A}(\rho)$ and similarly we have $\rho_{A}= \Tr_{B}(\rho)$ obtained from the full density matrix after tracing over subsystem $B$. The ensemble of the reduced density matrix is the trace restricted Wishart ensemble (trace being unity).

The average density of states of ensemble of Wishart matrices is given by the Marchenko--Pastur distribution \cite{mehta2004random} 
\begin{align}
    f(\epsilon)&= \frac{\mathcal{N} Q}{2 \pi} \frac{\sqrt{(\epsilon_\mathrm{max}-\epsilon)(\epsilon-\epsilon_\mathrm{min})}}{\epsilon},\nonumber \\
    \epsilon^\mathrm{max}_\mathrm{min} &= \frac{1}{\mathcal{N}}\left(1+\frac{1}{Q} \pm \frac{2}{\sqrt{Q}}\right)
\end{align}
where $\epsilon \in [\epsilon_\mathrm{min},\epsilon_\mathrm{max}]$ and $Q= \mathcal{M}/\mathcal{N}$.

Using this, in \cite{Bandyopadhyay_2002} a bound on the average entanglement entropy was proposed, which is
\begin{equation}\label{eq:ee3f2}
    \braket{S_E} \cong  - \int_{\epsilon_\mathrm{min}}^{\epsilon_\mathrm{max}} f(\epsilon) \epsilon  \ln(\epsilon) d\epsilon \equiv \ln(\gamma \mathcal{N}).
\end{equation}

The von Neumann entropy is then given by 
\begin{equation}\label{eqn:ee}
   S( \rho_{A})= -\Tr( \rho_{A} \ln(\rho_{A})) = - \sum_{i = 1}^{\mathcal{N}} \epsilon_{i} \ln(\epsilon_{i})
\end{equation}
and similarly for $\rho_{B}$.
We will now give an alternative derivation of the integral given by Eq. \eqref{eq:ee3f2}.

We instead consider the following integral which is closely related to Eq. \eqref{eq:ee3f2} 
\begin{align}\label{eq:newint}
    I = \int_{a}^{b} \sqrt{(x-a)(b-a)} \ln(x) dx .
\end{align}
The integral is not straightforward to evaluate in the present form due to the presence of the logarithm function. We try to simplify the integral using the following 
\begin{align}
    \frac{d}{dx}(x^{\epsilon}) = x^{\epsilon} \ln(x) \nonumber \\
 \implies \frac{d}{dx}(x^{\epsilon})\Big|_{\epsilon \rightarrow 0} = \ln(x)
\end{align}
We thus re-write the integral in Eq. \eqref{eq:newint}
\begin{align}
I =     \frac{d}{dx} \left(\int_{a}^{b} \sqrt{(x-a)(b-x)} x^{\epsilon} dx \right)\Bigg|_{\epsilon \rightarrow 0}
\end{align}

We now have a simpler integral to evaluate. The above integral can be evaluated in terms of hypergeometric function (using \textsc{Mathematica}) and we obtain
\begin{align}
\int_{a}^{b} \sqrt{(x-a)(b-a)} x^{\epsilon} dx =    \frac{a^{\epsilon -1} \left(\pi  a (a+b) \, _2F_1\left(-\frac{1}{2},1-\epsilon ;1;1-\frac{b}{a}\right)-\pi  b (2 a \epsilon +a-2 b \epsilon +b) \, _2F_1\left(\frac{1}{2},1-\epsilon ;1;1-\frac{b}{a}\right)\right)}{4 \epsilon  (\epsilon +1)}
\end{align}

The next step to evaluate $I$ is to take the derivative of the above result and take the limit $\epsilon \rightarrow 0$. Using the integral representation of the $_2F_{1}$ hypergeometric integral we get the following
\begin{align}
    \, \frac{d}{dx}\left(_2F_1\left(\frac{1}{2},1-\epsilon ;1;1-\frac{b}{a}\right) \right)\Bigg|_{\epsilon=0}&= 2 \log \left(\frac{2}{\sqrt{\frac{b}{a}}+1}\right), \\
\frac{d}{dx}\left( _2F_1\left(-\frac{1}{2},1-\epsilon ;1;1-\frac{b}{a}\right)\right)\Bigg|_{\epsilon=0} &=2 \left(-1+\log \left(\frac{2}{\sqrt{\frac{b}{a}}+1}\right)+\sqrt{\frac{b}{a}}
 \right).
\end{align}
Using these results we obtain the result of integral in Eq. \eqref{eq:newint} as follows
\begin{equation}
  I= \frac{1}{16} \pi  \left(-4 a^2 \sqrt{\frac{b}{a}}+a^2+6 a b-4 a b \sqrt{\frac{b}{a}}-4 \log (2) (a-b)^2+2 (a-b)^2 \log \left(\left(\sqrt{a}+\sqrt{b}\right)^2\right)+b^2\right)
\end{equation}
It is now straightforward to evaluate the integral in Eq. \eqref{eq:ee3f2}. We get a remarkably simpler result for $I$ as compared to the result previously stated (in \cite{Bandyopadhyay_2002}).
\begin{equation}
    S= \ln(\mathcal{N}) -\frac{1}{2 Q} = \ln(\mathcal{N}) - \frac{\mathcal{N}}{2 \mathcal{M}}.
\end{equation}
We remark that the above result is not new but it is the Page's result \cite{Page:1993df} for the entanglement entropy of a random pure state in the limit $Q \gg1$.

\clearpage

\section{More results for SYK model}
In this section we present results for different types of SYK model for various other value of sparseness, which are not shown in the main text. 

\subsection{The SYK model}
\begin{figure}[htbp]
    \centering
    \includegraphics[width=  
    0.85\linewidth]{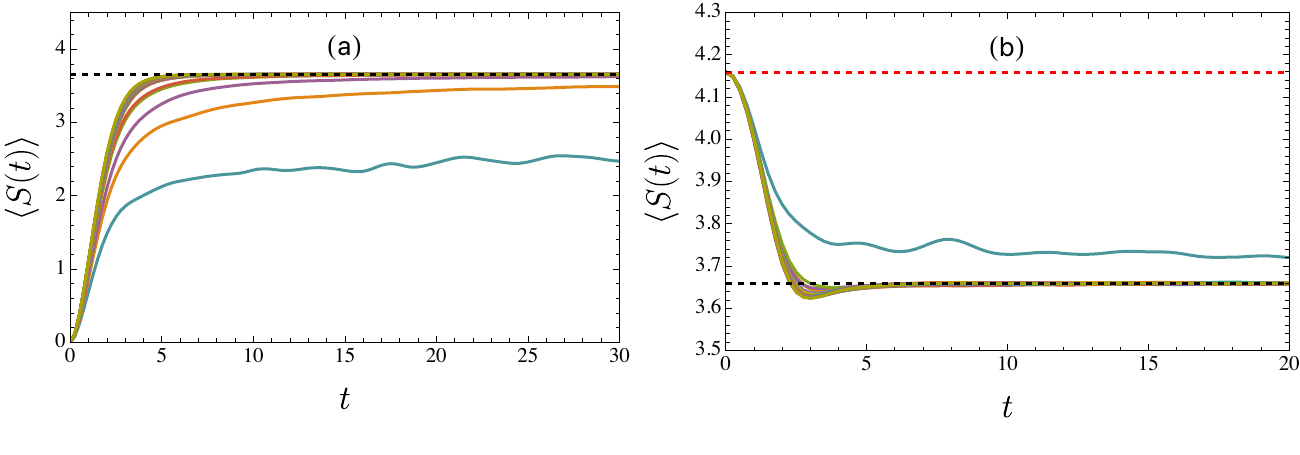}
      \caption{(a) EE evolution of a product state with time, for various system fractions. The horizontal black line corresponds to the analytical result given by Eq. \eqref{eq:eesimplesupp}. We take $N =24$ and averaging is done over 50 realizations. We take values as $p$= \textcolor[RGB]{94, 129, 181}{0.001},\textcolor[RGB]{225, 156, 36}{0.002},\textcolor[RGB]{143, 176, 50}{0.003},\textcolor[RGB]{235, 98, 53}{0.004},\textcolor[RGB]{135, 120, 179}{0.005},\textcolor[RGB]{197, 110, 26}{0.01},\textcolor[RGB]{93, 158, 199}{0.012},\textcolor[RGB]{255, 191, 0}{0.013},
\textcolor[RGB]{165, 96, 157}{0.014},\textcolor[RGB]{146, 150, 0}{0.015},\textcolor[RGB]{233, 85, 54}{0.016},\textcolor[RGB]{102, 133, 217}{0.017},\textcolor[RGB]{248, 159, 19}{0.018},\textcolor[RGB]{188, 91, 128}{0.019},\textcolor[RGB]{71, 182, 109}{0.02},\textcolor[RGB]{214, 114, 5}{0.03},\textcolor[RGB]{149, 105, 211}{0.04},\textcolor[RGB]{229, 190, 0}{0.05},\textcolor[RGB]{215, 88, 84}{0.06},
\textcolor[RGB]{72, 155, 192}{0.07},\textcolor[RGB]{238, 135, 24}{0.08},\textcolor[RGB]{172, 92, 153}{0.1},\textcolor[RGB]{138, 182, 25}{0.25},\textcolor[RGB]{226, 96, 36}{0.5},\textcolor[RGB]{121, 122, 217}{1}. (b) EE evolution of a maximally entangled state with time, for various system fractions. The horizontal black line corresponds to the analytical result given by Eq. \eqref{eq:eesimplesupp}. We take  $N =24$ and averaging is done over 50 realizations. The critical value of $p$ after which the the entanglement entropy does not reach the theoretical bound, for initial product state, is $p_{c}= 0.004$. For the case when initial state is a maximally  entangled state $p_{c}= 0.001$ }
\end{figure}

\subsection{The Binary SYK model}
\begin{figure}[htbp]
    \centering
    \includegraphics[width=  0.85\linewidth]{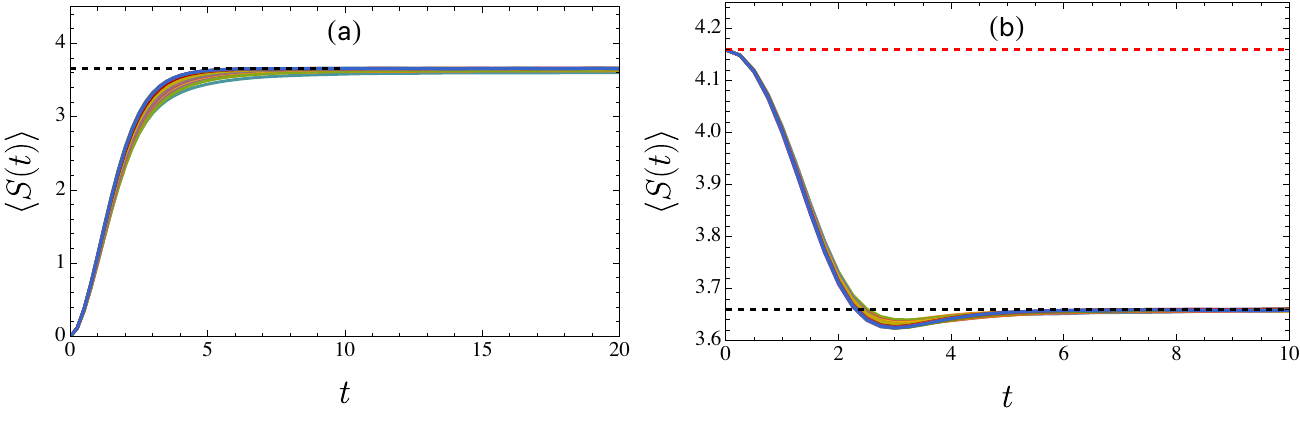}
    \caption{(a) EE evolution of a product state with time, for various system fractions, for binary sparse SYK model. The horizontal black line corresponds to the analytical result given by Eq. \eqref{eq:eesimplesupp}. We take $N =24$ and averaging is done over 50 realizations. We take values as $\kappa$= \textcolor[RGB]{94, 129, 181}{18},\textcolor[RGB]{225, 156, 36}{20},\textcolor[RGB]{143, 176, 50}{22},\textcolor[RGB]{235, 98, 53}{24},\textcolor[RGB]{135, 120, 179}{26},\textcolor[RGB]{197, 110, 26}{28},\textcolor[RGB]{93, 158, 199}{30},\textcolor[RGB]{255, 191, 0}{32},\textcolor[RGB]{165, 96, 157}{34},\textcolor[RGB]{146, 150, 0}{36},\textcolor[RGB]{233, 85, 54}{38},\textcolor[RGB]{102, 133, 217}{40},\textcolor[RGB]{248, 159, 19}{100},\textcolor[RGB]{188, 91, 128}{600},\textcolor[RGB]{71, 182, 109}{900},\textcolor[RGB]{214, 114, 5}{1500},\textcolor[RGB]{149, 105, 211}{3500},\textcolor[RGB]{229, 190, 0}{6500},\textcolor[RGB]{215, 88, 84}{10626}. (b) EE evolution of a maximally entangled state with time, for various system fractions. The horizontal black line corresponds to the analytical result given by Eq. \eqref{eq:eesimplesupp}. We take  $N =24$ and averaging is done over 50 realizations. Observe how even for $\kappa= 18$ ($p \approx 0.001$ the curves are still very close to the $\kappa= 10626$ curve, for both kinds of initial states. The critical value of $\kappa$ after which the the entanglement entropy does not reach the theoretical bound is $k_{c}= 22$ or $p_{c}= 0.002$. For the case when initial state is a maximally entangled state we find that even for the smallest $\kappa$ (=18) the bound is saturated.}
\end{figure}
\clearpage

\subsection{The Spin-SYK model}

\begin{figure}[htbp]
    \centering
    \includegraphics[width=  0.85\linewidth]{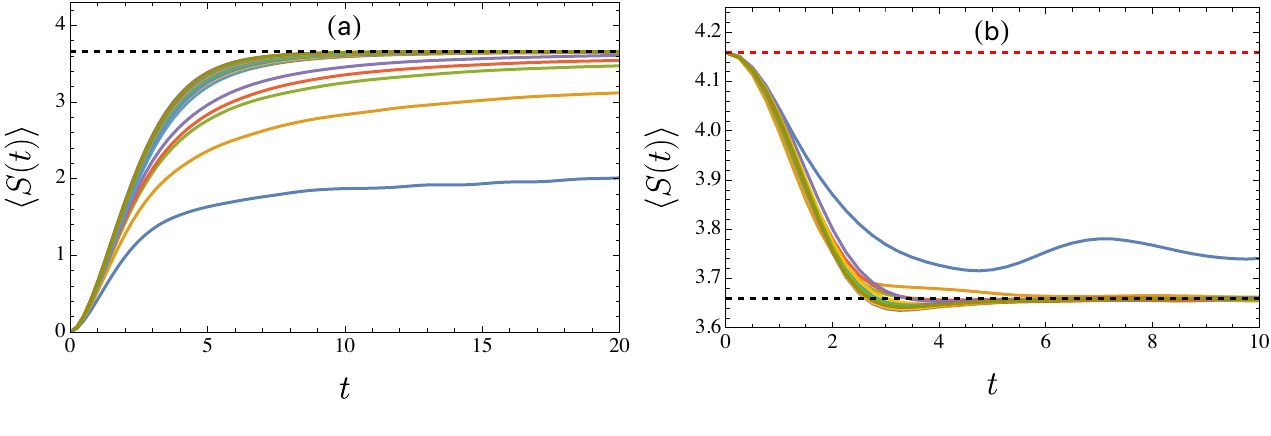}
    \caption{(a) EE evolution of an initial product state with time, for various system fractions, for the sparse spin SYK model. The horizontal black line corresponds to the analytical result given by Eq. \eqref{eq:eesimplesupp}. We take $N =12$ and averaging is done over 50 realizations. We take values as $p$= \textcolor[RGB]{94, 129, 181}{0.001},\textcolor[RGB]{225, 156, 36}{0.002},\textcolor[RGB]{143, 176, 50}{0.003},\textcolor[RGB]{235, 98, 53}{0.004},\textcolor[RGB]{135, 120, 179}{0.005},\textcolor[RGB]{197, 110, 26}{0.01},\textcolor[RGB]{93, 158, 199}{0.012},\textcolor[RGB]{255, 191, 0}{0.013},
\textcolor[RGB]{165, 96, 157}{0.014},\textcolor[RGB]{146, 150, 0}{0.015},\textcolor[RGB]{233, 85, 54}{0.016},\textcolor[RGB]{102, 133, 217}{0.017},\textcolor[RGB]{248, 159, 19}{0.018},\textcolor[RGB]{188, 91, 128}{0.019},\textcolor[RGB]{71, 182, 109}{0.02},\textcolor[RGB]{214, 114, 5}{0.03},\textcolor[RGB]{149, 105, 211}{0.04},\textcolor[RGB]{229, 190, 0}{0.05},\textcolor[RGB]{215, 88, 84}{0.06},
\textcolor[RGB]{72, 155, 192}{0.07},\textcolor[RGB]{238, 135, 24}{0.08},\textcolor[RGB]{172, 92, 153}{0.1},\textcolor[RGB]{138, 182, 25}{0.25},\textcolor[RGB]{226, 96, 36}{0.5},\textcolor[RGB]{121, 122, 217}{1}. (b) EE evolution of a maximally entangled state with time, for various system fractions. The horizontal black line corresponds to the analytical result given by Eq. \eqref{eq:eesimplesupp}. We take  $N =12$ and averaging is done over 50 realizations. The critical value of $p$ after which the the entanglement entropy does not reach the theoretical bound, for initial product state, is $p_{c}= 0.01$, which is much higher as compared to SYK$_b$ and SYK model. For the case when initial state is a maximally entangled state we find $p_{c}= 0.001$, comparable to SYK model.}
\end{figure}
\subsection{Autocorrelation function}

We now show further supporting results for the behavior of autocorrelation function.

\begin{figure}[h]
    \centering
    \includegraphics[width=  \linewidth]{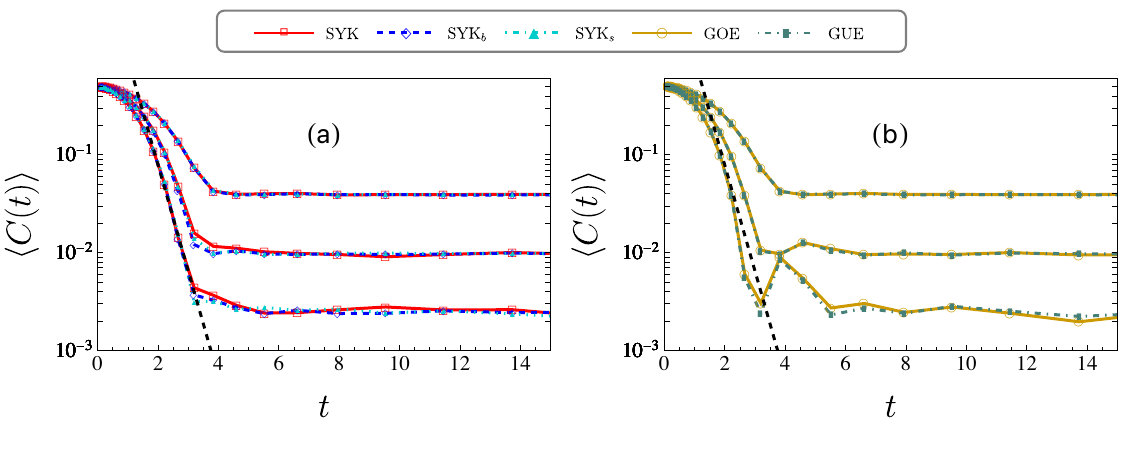}
    \caption{(a) Autocorrelation function for Majorana operator $\hat{O}= \psi_{1}$, for different kinds of SYK model. The overlapping legends are for SYK, SYK$_b$ and SYK$_s$ for $N= 16
    (10000), 24(500), 32(50)$ of Majorana fermions (and $N/2$ number of spins for SYK$_s$). The number in the bracket denotes the number of Hamiltonian realizations. We only consider odd parity sector. (b) Autocorrelation function for corresponding random matrices of GOE and GUE type of dimensions $2^{N/2-1}$. Black dashed curve represent the best fit curve; $a e^{-b x} $; $b= 2.49 \pm 0.23 $ ,  in the exponential decay regime obtained using the data for $2^{15}$ dimensional random matrices.}
    \label{fig:acfalls}
\end{figure}

\begin{figure}[h]
    \centering
    \includegraphics[width=  \linewidth]{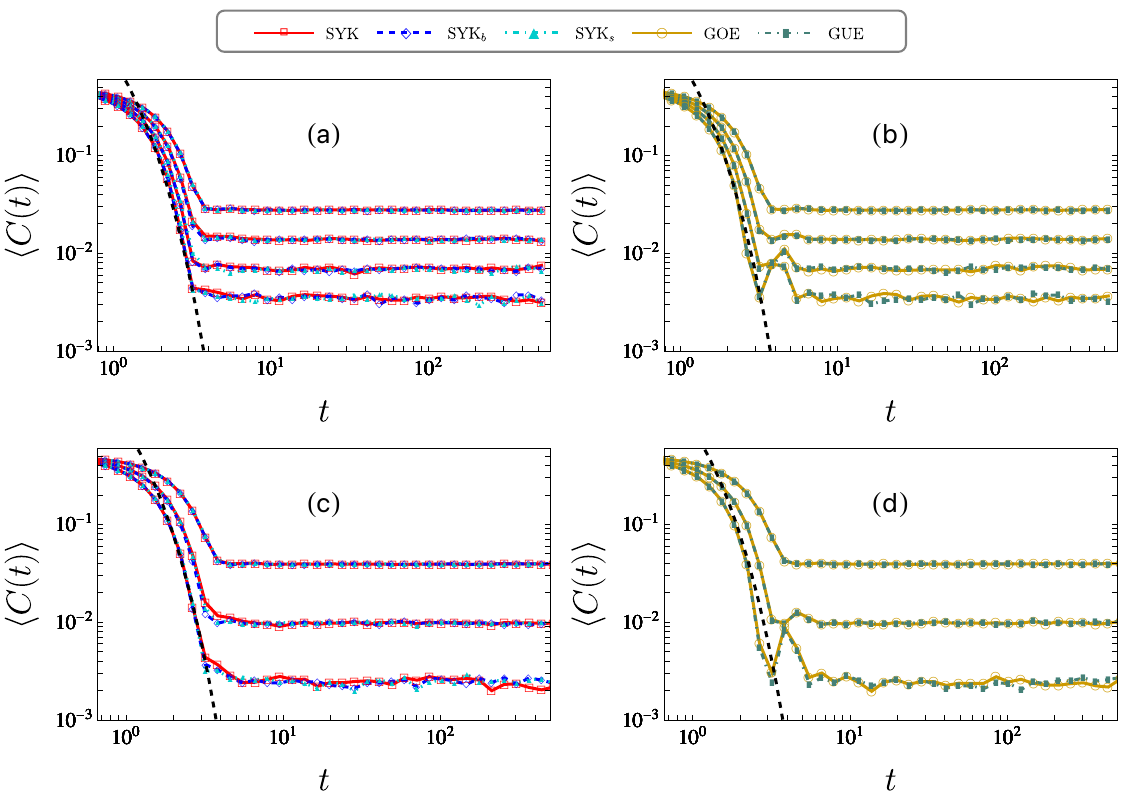}
    \caption{(a) Late time behavior fo the autocorrelation function (in a ln-ln plot) for Majorana operator $\hat{O}= \psi_{1}$, for different kinds of SYK model. The overlapping legends are for SYK, SYK$_b$ and SYK$_s$ for $N= 18 (5000), 22 (1000), 26(200),30(100)$ of Majorana fermions (and $N/2$ number of spins for SYK$_s$). The number in the bracket denotes the number of Hamiltonian realizations. We only consider odd parity sector. (b) Autocorrelation function for corresponding random matrices of GOE and GUE type of dimensions $2^{N/2-1}$. Black dashed curve represent the best fit curve; $a e^{-b x} $; $b= 2.49 \pm 0.23 $ ,  in the exponential decay regime obtained using the data for $2^{14}$ dimensional random matrices. (c) and (d) Same as (a) and (b) but for $N= 16
    (10000), 24(500), 32(50)$ of Majorana fermions (and $N/2$ number of spins for SYK$_s$) and random matrices of dimension $2^{N/2-1}$.}
    \label{fig:acfallslate}
\end{figure}
\bibliography{references}